\documentclass[12pt]{article}
\usepackage{scicite}
\usepackage{times}
\usepackage{graphicx}

\newenvironment{sciabstract}{%
\begin{quote} \bf }
{\end{quote}}

\newcounter{lastnote}
\newenvironment{scilastnote}{%
\setcounter{lastnote}{\value{enumiv}}%
\addtocounter{lastnote}{+1}%
\begin{list}%
{\arabic{lastnote}.}
{\setlength{\leftmargin}{.22in}}
{\setlength{\labelsep}{.5em}}}
{\end{list}}

\title{
Metal-Insulator Transition in  Disordered Two-Dimensional Electron Systems.
}
\author
{Alexander Punnoose,$^{1\ast}$ Alexander M Finkel'stein$^{2}$\\
\\
\normalsize{$^{1}$Bell Laboratories, Lucent Technologies,  Murray Hill, NJ 07974, USA.}\\
\normalsize{$^{2}$Department of Condensed Matter Physics, Weizmann Institute of
Science, Rehovot 76100, Israel.}\\
\\
\normalsize{$^\ast$To whom correspondence should be addressed; E-mail:  punnoose@lucent.com.}
}

\date{}

\begin{document}
\baselineskip24pt

\maketitle

\begin{sciabstract}
We present a theory of the metal-insulator transition in a
disordered two-dimensional electron gas. A quantum critical point,
separating the metallic phase which is stabilized by electronic
interactions, from the insulating phase where disorder prevails over
the electronic interactions, has been identified.
The existence of  the quantum critical point leads to a divergence in the density of states of the underlying collective modes at the transition, causing the thermodynamic
properties to behave critically as the transition is approached. We show that the interplay of electron-electron interactions and disorder
can explain the observed transport properties
and the anomalous enhancement of the spin susceptibility  near the metal-insulator transition.
\end{sciabstract}

\newpage
Measurements in high-mobility two-dimensional (2D) semiconductors  have made it possible to study the properties of the 2D electron gas at low carrier densities. In a clean system at very low carrier densities, the electrons are expected to solidify into a Wigner crystal. In a wide range of low electron concentrations, prior to the Wigner crystal phase, the electron system exists as a strongly interacting liquid~\cite{ceperely}. Understanding the properties of this strongly interacting system in the presence of disorder has proven to be an extremely challenging theoretical problem. The recent experimental discovery of a metal-insulator transition (MIT) in 2D~\cite{krav94}, when none was expected for more than a decade~\cite{gang4}, generated renewed interest in this field~\cite{kravreview04}.

The MIT, observed initially in high-mobility silicon metal-oxide-semiconductor field-effect transistors (Si-MOSFETs) and subsequently in a host of other 2D systems like GaAs heterostructures, occurs when the resistance $R$  is of the order of the quantum resistance $h/e^2$, emphasizing the importance of quantum effects.  In the metallic phase the resistance drops noticeably as the temperature is lowered. This drop is  suppressed when  an in-plane magnetic field is applied~\cite{kravreview04}. Additionally, Si-MOSFET samples show a strong enhancement of the spin susceptibility as the MIT is approached~\cite{mishachi03,kravchi04}. There are indications  that the spin susceptibility in  samples of different mobilities behaves critically near the transition. The sensitivity to in-plane magnetic fields together with the anomalous behavior of the spin susceptibility highlight the importance of electron-electron ($e\textrm{-}e$) interactions in this phenomenon.

In the presence of impurities, perturbations of the charge and
spin densities (if spin is conserved) relax diffusively at low frequencies and large distances. Formally, this relaxation occurs via the propagation of particle-hole pair modes. In a system which obeys time-reversal symmetry, the modes in the particle-particle channel (i.e., the Cooper channel) also have a diffusive form, leading to the celebrated weak-localization corrections to the conductivity. These two mode families are, in conventional terminology, referred to as Diffusons and Cooperons, respectively. Taken together, they describe the low energy dynamics of a disordered electron liquid~\cite{AAbook}. The modes have the diffusive form
$\mathcal{D}(q,\omega)=1/(Dq^2-iz\omega)$, where $D$  is the diffusion coefficient, and the parameter  $z$  determines the relative scaling of the frequency (i.e., energy) with respect to the length scale~\cite{sasha83,sasha83b}; $z=1$  for free electrons. The addition of $e\textrm{-}e$  interactions results in the scattering of these diffusion modes. Both $D$  and $z$  therefore, acquire corrections in the presence of the interactions. Conversely, diffusing electrons dwell long in each other's vicinity, becoming more correlated at low enough energies. As a result, the $e\textrm{-}e$  scattering amplitudes $\gamma_2$  and $\gamma_c$  characterizing the scattering of the Diffuson and Cooperon modes, respectively, acquire corrections as a function of the disorder. All these corrections, due to the diffusive form of the propagator $\mathcal{D}(q,\omega)$, are logarithmically divergent in temperature in 2D~\cite{AAL}. They signal the breakdown of perturbation theory and the need for a re-summation.

A consistent handling of these mutually coupled corrections to $D$ and $z$ on the one hand, and $\gamma_2$ and $\gamma_c$ on the other, requires the use of renormalization group (RG) that effectively sums the logarithmic series~\cite{sasha83,sasha84,castellani84}, allowing one to approach the strong-coupling region of the MIT~\cite{sasha83b,pruisken02}. We show that an internally consistent solution of the MIT
can be obtained within a suitably defined large$\textrm{-}N$ model involving $N$ flavors of electrons.
Bearing  in mind that  the conduction band in semiconductors often has almost degenerate regions
called valleys~\cite{valleys}, the electrons carry both  spin and  valley indices. Taking the number of valleys $n_v\rightarrow \infty$, we derive the
relevant RG equations to two-loop order, and show that a quantum critical point (QCP) that describes the
MIT exists in the 2D interacting disordered electron liquid.

As the inter-valley scattering requires a large change of the momentum, we assume that the interactions couple electrons in different valleys but do not mix them, i.e., inter-valley scattering processes, including those due to the disorder, are neglected. This assumption is appropriate for samples with high-mobility. In this limit, the RG equations describing the evolution of the resistance and the scattering amplitude $\gamma_2$  in 2D are known to have the form~\cite{punnoose02}:
%
\begin{eqnarray}
\frac{d\ln\rho}{d\xi}&=&\rho\;\left[n_v+1-(4n_v^2-1)\left(\frac{1+\gamma_2}{\gamma_2}\ln(1+\gamma_2)-1\right) \right],
\label{eqn:onelooprho}\\
\frac{d\gamma_2}{d\xi}&=&\rho\;\frac{(1+\gamma_2)^2}{2},
\label{eqn:oneloopgamma2}
\end{eqnarray}
%
where $\xi=-\ln(T\tau/\hbar)$ with $T\tau\ll \hbar$ (diffusive regime), $\tau$ is
the elastic scattering time and the dimensionless resistance parameter $\rho=(e^2/\pi h)R$
is related to $D$ as $\rho=1/(2\pi)^2 n_v \nu D$, with $\nu$ being the density of
states of a single spin and valley species. (For repulsive interactions, the scattering amplitude $\gamma_c$ scales to zero when $n_v$  is finite and is therefore, neglected for now; the situation in the large$\textrm{-}n_v$  limit is different and is discussed later.) Though the initial values of $\rho$  and $z$, determined at some initial temperature, depend on the system and are therefore not universal, the flow of $\rho$  can be described for each $n_v$
by a universal function $R(\eta)$~\cite{sasha83}:
\begin{equation}
\rho=\rho_{\mathrm{max}} R(\eta)\hspace{0.25cm}\mathrm{and}\hspace{0.25cm}\eta=\rho_{\mathrm{max}}\ln(T_{\mathrm{max}}/T)~,
\label{eqn:singlecurve}
\end{equation}
where $R(\eta)$  is a non-monotonic function with a maximum at a temperature $T_{\mathrm{max}}$  corresponding to the point where $d\rho/d\xi$  in Eq.~(\ref{eqn:onelooprho}) changes sign. It follows that the full temperature dependence of the resistance $\rho$  is  completely controlled by its value $\rho_{\mathrm{max}}$  at the maximum; there are no other free (or fitting) parameters. Note that the anti-localization effect of the $e\textrm{-}e$ interactions fundamentally alters the commonly accepted point of view~\cite{gang4} that in 2D all states are ``eventually" (i.e., at $T=0$) localized.

This solution has obvious limitations, however. Because the RG equations were derived in the lowest order in $\rho$, the single curve solution $R(\eta)$  in Eq.~(\ref{eqn:singlecurve}) cannot be applied in the critical region of the MIT where $\rho\sim 1$. (In fact, for $n_v=2$, $R(\eta)$  describes quantitatively the temperature dependence of the resistance of high-mobility Si-MOSFETs  in the region of $\rho$  up to $\rho\sim 0.5$~\cite{punnoose02}, which is not so far from the critical region.) Therefore, to approach the MIT the disorder has to be treated beyond the lowest order in $\rho$,  while adequately retaining the effects of the interaction. This also touches upon the delicate issue of the internal consistency and nature of the theory as $T\rightarrow 0$. The problematic feature of the scaling Eqs.~(\ref{eqn:onelooprho}) and (\ref{eqn:oneloopgamma2}) is that the amplitude $\gamma_2$  diverges at a finite temperature $T^*$  and thereafter the RG theory becomes uncontrolled~\cite{sasha84,castellani84}. Fortunately, the scale $T^*$ decreases very rapidly with $n_v$  as $\ln\ln(1/\tau T^*)\sim (2n_v)^2$, making the problem of the divergence of $\gamma_2$  for all practical purposes irrelevant even for $n_v=2$. Still, to get an internally consistent solution up to $T=0$  it is useful to study the limit $n_v\rightarrow\infty$, for which $T^*\rightarrow 0$.

The valley degrees of freedom are akin to flavors in standard field-theoretic models. Generally, closed loops play a special role in the diagrammatic RG analysis in the limit when the number of flavors $N$ is taken to be very large~\cite{wilson73}. This is because each closed loop involves a sum over all the flavors,  generating a large factor $N$  per loop. It is then typical to send a coupling constant $\lambda$  to zero in the limit $N\rightarrow\infty$ keeping $\lambda N$ finite. For interacting spin-$1/2$ electrons in the presence of $n_v$ valleys ($N=2n_v$), the screening being enhanced makes the bare values of the electronic interaction amplitudes $\gamma_2$  and $\gamma_c$  to scale as $1/2n_v$. Furthermore, the increase in the number of conducting channels results in the resistance $\rho$  to scale as $1/n_v$. It is therefore natural to introduce the amplitudes $\theta_2=2n_v\gamma_2$  and $\theta_c=2n_v\gamma_c$  together with the resistance parameter $t=n_v\rho$, which remain finite in the large$\textrm{-}n_v$ limit. The parameter  $t$ is the resistance per valley, $t=1/(2\pi)^2\nu D$, and it reveals itself in the theory via the momentum integration involving the diffusion propagators  $\mathcal{D}(q,\omega)$.  In terms of these variables, the contribution of a closed loop connected to the rest of the diagram by one interaction amplitude, $\gamma_2$  (or $\gamma_c$), after integrating over the momentum flowing through the loop, and summing over the spin and valley degrees of freedom, is proportional to $2n_vt\gamma_2=t\theta_2$  (or  $t\theta_c$). While, this contribution remains finite at large$\textrm{-}n_v$, those diagrams with more than one interaction for every closed loop are negligible. This one-to-one correspondence between the number of loops in a given diagram and the number of interaction amplitudes limits the  maximum number of interaction vertices for a given power of $t$. This is the crucial simplification on taking the large$\textrm{-}n_v$  limit.

For repulsive interactions at finite$\textrm{-}n_v$, re-scattering in the Cooper channel leads to the vanishing of the effective amplitude $\gamma_c$ at low energies. The amplitude $\theta_c$ is, however, relevant in the large$\textrm{-}n_v$ limit as the re-scattering is not accompanied by  factors of $n_v$,
We introduce the parameter $\alpha=1$  to mark the contributions arising from the Cooper channel. Violation of time reversal symmetry (e.g., by a magnetic field) suppresses the Cooperon modes~\cite{AAbook}. Setting $\alpha=0$  switches the Cooper channel off.
Following the large$\textrm{-}n_v$  approximation scheme detailed above, the RG equations to order $t^2$  are derived for $\alpha=0$  and  $1$:
%
\begin{eqnarray}
\frac{d\ln t}{d\xi}&=&t\left(\alpha-\Theta\right)+t^2(1-\alpha-\alpha\Theta+c_t\Theta^2),
\label{eqn:twoloopt}\\
\frac{d\Theta}{d\xi}&=&t\left(1+\alpha+\alpha\Theta\right)-
4t^2\left((1+\alpha)\Theta+\frac{\alpha}{2}\Theta^2+c_{\theta}\Theta^3\right).
\label{eqn:twolooptheta}
\end{eqnarray}
%
The constants  $c_t=(5-\pi^2/3)/2\approx 0.8$ and $c_{\theta}=(1-\pi^2/12)/2\approx 0.08$.
It turns out that in the  large$\textrm{-}n_{v}$ limit  the
amplitudes $\theta_2$ and $\theta_c$  appear together in the combination
$\Theta=\theta_2+\alpha\theta_c$.  The fact that they come together as a single parameter $\Theta$  is unique to the large$\textrm{-}n_v$  limit.  Eq.~(\ref{eqn:twoloopt}) reproduces the known result to order $t^2$ when the
electronic interactions are absent~\cite{efetov82}. Notice that while the maximum power of $\Theta$  is limited by the order of $t$, the opposite is not true; it is the number of momentum integrations of the diffusion propagators that determine the power of $t$.

\begin{figure}
\centerline{\includegraphics[width=0.45\textwidth]{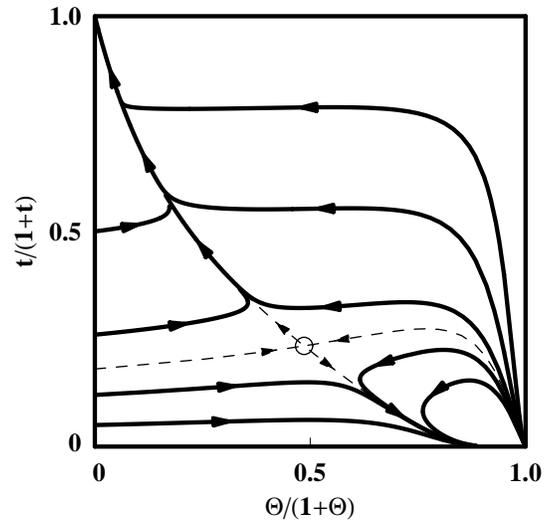}}
\caption{The disorder-interaction, $t\textrm{-}\Theta$, flow diagram of the 2D electron gas obtained by solving Eqs.~(\ref{eqn:twoloopt}) and (\ref{eqn:twolooptheta}) with the  Cooper channel included ($\alpha=1$).  Arrows indicate the direction of the flow as the temperature is lowered. The circle denotes the QCP of the MIT, and the dashed lines show the separatrices.}
\label{fig:flowdiag}
\end{figure}

Eqs.~(\ref{eqn:twoloopt}) and (\ref{eqn:twolooptheta}) describe the competition
between the electronic interactions and disorder in 2D.
The flow of $t(\xi)$ and $\Theta(\xi)$
are plotted in Fig.~\ref{fig:flowdiag} for the case $\alpha=1$ (the flows are qualitatively
the same for $\alpha=0$). The arrows indicate the direction of the flow as the temperature is lowered. The QCP, corresponding to the fixed point  of Eqs.~(\ref{eqn:twoloopt}) and (\ref{eqn:twolooptheta}), is marked by the circle. The attractive (``horizontal") separatrices separate the
metallic phase where $t\rightarrow 0$ from the insulating phase where $t\rightarrow \infty$.
Crossing one of these separatrices~\cite{whichseparatrix} by changing the initial values of $t$ and $\Theta$ (e.g., by changing the carrier density)
leads to the MIT. It is interesting that near the fixed point these separatrices
are  almost insensitive to temperature, which is in agreement with the experiments in Si-MOSFETs~\cite{klapwijk00,fig3}.
The accidental (but fortunate) smallness of the fixed point parameters  $t_c\approx 0.3$ and  $t_c\Theta_c\approx 0.27$  permits us to believe that the two-loop equations derived in the large$\textrm{-}n_v$  limit capture accurately enough the main features of the transition.

We now discuss the renormalization of the parameter $z$ due to the $e\textrm{-}e$ interactions.
The RG equation for $z$ is described by an independent equation, which quite generally~\cite{sasha84} takes the form $d\ln z/d\xi=\beta_z(t,\gamma_2,\gamma_c; n_v)$.
Observe that $\beta_z$, as well as  Eqs.~(\ref{eqn:twoloopt}) and (\ref{eqn:twolooptheta}) for $t$ and $\Theta$, are all independent of $z$. Consequently, $z$  in the vicinity of the MIT is critical, i.e., $z\sim 1/T^{\zeta}$ with a critical exponent  $\zeta$ equal to the value of the function $\beta_z$ at the critical point~\cite{sasha83b,pruisken02}.
The parameter $z$, being related to the frequency renormalization  can be interpreted as the density of states of the diffusion modes, and therefore controls the contribution of the diffusion modes to the specific heat, i.e., $C_V\sim (z\nu) T$~\cite{castellani86,castellani87}. Hence, in the critical regime $C_V\sim T^{1-\zeta}$.  (The fact that $C_V$  must vanish as $T\rightarrow 0$  constraints the exponent $\zeta<1$.) Furthermore, the Pauli spin susceptibility $\chi$ in the disordered electron liquid has the general form $\chi\sim (z\nu)(1+\gamma_2)$~\cite{sasha84,castellani84}, i.e., apart from the Stoner-like enhancement of the  $g$-factor, $(1+\gamma_2)$, it is proportional to the renormalized density of states $z$.  The enhancement of $z$ near the QCP implies that the spin susceptibility diverges as $\chi\sim 1/T^{\zeta}$, while the spin diffusion coefficient $D_s=D/z(1+\gamma_2)$  scales to zero. Note that at the fixed point the  $g$-factor remains finite and therefore the divergence of the spin susceptibility is a priori not related to any magnetic instability.
In the large$\textrm{-}n_v$  limit, we find that no terms of the order $t^2$  are  generated in the equation for~$z$:
\begin{equation}
\frac{d\ln z}{d\xi}=t\Theta~.
\label{eqn:twoloopz}
\end{equation}
Therefore, we get for the  critical exponent  $\zeta\approx 0.27$, which is noticeably smaller than one.

We propose that the existence of the QCP explains the anomalous enhancement of the spin
susceptibility that has been  observed near the MIT in Si-MOSFETs with different critical densities~\cite{mishachi03,kravchi04}. Additionally, an important consequence of this theory
is that the compressibility
\protect{$\partial n/\partial \mu$} is regular across the transition, in agreement
with the experiments in  p-GaAs~\cite{jiang00,yacoby01}.

In the insulating phase the parameter $t$  diverges at a finite scale $\xi_c$, indicating the onset of strong localization. Since the  $t^2$ term in Eq.~(\ref{eqn:twolooptheta}) is negative, it forces the interaction amplitude $\Theta$  to vanish in the insulating phase. It can be shown  that the parameter $z$  is finite at the scale $\xi_c$. The vanishing of the interaction amplitude $\Theta$  and a finite value of $z$  makes the insulating phase to be similar to the  Anderson insulator.

On the metallic side, a state with decreasing resistance $t\rightarrow 0$ and enhanced amplitude $\Theta\rightarrow\infty$ is stabilized  by the electronic interactions as $T\rightarrow 0$  in the large$\textrm{-}n_v$  limit.
It can be shown within this solution that $z\sim 1/t$,  implying that the quasi-particle diffusion coefficient defined as $D_{{qp}}=D/z$~\cite{castellani87} behaves regularly in the metallic phase.
Deep in the metallic phase, the enhancement in $\chi$  will be observed only at exponentially small temperatures.
This holds even in the case of finite$\textrm{-}n_v$, because the scale $T^*$ at which the RG Eqs.~(\ref{eqn:onelooprho}) and (\ref{eqn:oneloopgamma2}) become uncontrolled is immeasurably small even for $n_v=2$.
Note that the question  of the existence of the QCP at finite$\textrm{-}n_v$ is distinct from the problem of the divergence of the parameter $\gamma_2$ at $T=T^*$,  as
the two phenomena occur in different parts of the phase diagram.

A description of the MIT is obtained within the two-loop approximation in the limit of a large number of degenerate valleys. Although the  properties of the thermodynamic quantities in the critical region are obtained in the  large$\textrm{-}n_v$ limit,  it captures the generic features of the MIT for any $n_v$. Our solution gives a physical picture that is in qualitative agreement with the experimental situation. In particular, it is shown that the point of the MIT is accompanied by a divergence in the spin susceptibility whose origin is not related to any magnetic instability.

\newpage


\begin{thebibliography}{10}
\bibitem{valleys}
For example, the electron band in the $(001)$ plane in Si-MOSFET has
two
  valleys.
\bibitem{whichseparatrix}
The important question of which of the attractive separatrices are
crossed at
  the transition can be resolved by magneto-resistance measurements in a weak
  in-plane magnetic field.
\end{thebibliography}

\bibliographystyle{Science}


\begin{scilastnote}
\item AP benefited from discussions with Rob Whitney. AF was supported by the  MINERVA Foundation.
\end{scilastnote}

\newpage

\end{document}